\documentclass[12pt]{article}
\usepackage[margin=1in]{geometry}
\usepackage{setspace}
\usepackage{hyperref}

\usepackage[round,numbers,sort&compress]{natbib}
\usepackage{setspace}

\usepackage{comment}
\usepackage{tikz}
\usetikzlibrary{arrows.meta, positioning, calc, decorations.pathreplacing}

\usepackage{graphicx}
\usepackage{amsmath}
\usepackage{amssymb}
\usepackage{orcidlink}
\usepackage{booktabs}
\usepackage{longtable}
\usepackage{threeparttable}
\usepackage{caption}
\usepackage{xcolor}
\usepackage{colortbl} 
\usepackage{multirow}
\usepackage{array}
\usepackage{placeins}
\usepackage{mathtools}

\newcommand{\indep}{\perp \!\!\! \perp}
\hypersetup{
    colorlinks=true,
    linkcolor=blue,   
    urlcolor=cyan,
    citecolor=blue
    }

\usepackage{pdfpages}

\title{Prior-Data Fitted Networks for Causal Inference: a Simulation Study with Real-World Scenarios}
\author{
  Francisco Mourao$^{1,2}$\orcidlink{0009-0009-4757-0688}\thanks{Corresponding author: \href{mailto:francisco.dematosaguasmourao@aphp.fr}{francisco.dematosaguasmourao@aphp.fr}}, David Hajage$^{1,3}$\orcidlink{0000-0002-8475-4090}, Daria Bystrova$^{1}$, \\ Bertrand Bouvarel$^{1,2}$, Nathanaël Lapidus$^{1,2}$\orcidlink{0000-0002-4837-1068}, Fabrice Carrat$^{1,2}$\orcidlink{0000-0002-8672-7918}, \\ Benjamin Glemain$^{1,2}$\orcidlink{0009-0008-4771-2001}\\
  \vspace{0.1em} \\
  \small{$^1$Sorbonne Université, Inserm, Institut Pierre-Louis d’épidémiologie et} \\ \small{de santé publique, F-75012 Paris, France} 
  \\ 
  \small$^2$Département de santé publique, Hôpital Saint-Antoine, \\ \small AP-HP. Sorbonne Université, Paris, France 
  \\
  \small$^3$Hôpital Pitié-Salpêtrière, Département de Santé Publique, \\ \small Centre de Pharmacoépidémiologie, Sorbonne Université, Paris, France.
}
\date{}

\begin{document}
\maketitle

%\received{Date}{0}{Year}
%\revised{Date}{0}{Year}
%\accepted{Date}{0}{Year}

%\editor{Associate Editor: Name}

%\abstract{
%\textbf{Motivation:} .\\
%\textbf{Results:} .\\
%\textbf{Availability:} .\\
%\textbf{Contact:} \href{name@email.com}{name@email.com}\\
%\textbf{Supplementary information:} Supplementary data are available at \textit{Journal Name}
%online.}
\begin{abstract}

Prior-Data Fitted Networks (PFNs) represent a paradigm shift in tabular data prediction. We present the principles of this new paradigm and evaluate two PFNs for estimating the average treatment effect (ATE) of a binary treatment on a binary outcome, using simulated clinical scenarios based on real-world data. We assessed TabPFN combined with causal inference procedures (g-computation and inverse probability of treatment weighting), and CausalPFN, a PFN that directly provides an ATE estimate with a credible interval. Confidence intervals for the TabPFN-based methods were derived using bootstrap resampling. We found that computation times for TabPFN were prohibitive for routine causal inference, particularly because of the need for bootstrapping to yield confidence intervals. Moreover, g-computation with TabPFN produced a highly biased estimator, partially corrected by fitting separate models for each treatment group (T-learner). CausalPFN, by contrast, was computationally efficient but exhibited poor coverage of its 95\% credible interval for the ATE, due to both estimation bias and inadequate uncertainty quantification. Beyond automating model specification, some PFN variants---like CausalPFN---attempt to automate causal modeling. In the settings we evaluated, CausalPFN performed poorly. However, new algorithms of this kind continue to be developed, and their application to causal inference tasks requires further investigation.

\end{abstract}

\textbf{Prior-Data Fitted Networks, Causal Inference, TabPFN, G-Computation, Inverse Probability of Treatment Weighting, Automated Science}

\doublespacing

\maketitle

\newpage

\section*{Introduction}

In 2022, Müller et al. introduced Prior-Data Fitted Networks (PFNs), a new class of trans\-for\-mer-based deep learning algorithms~\cite{muller2022transformers}. PFNs represent a paradigm shift in tabular data prediction, as they are trained by their developers (prior to routine use), like large language models.

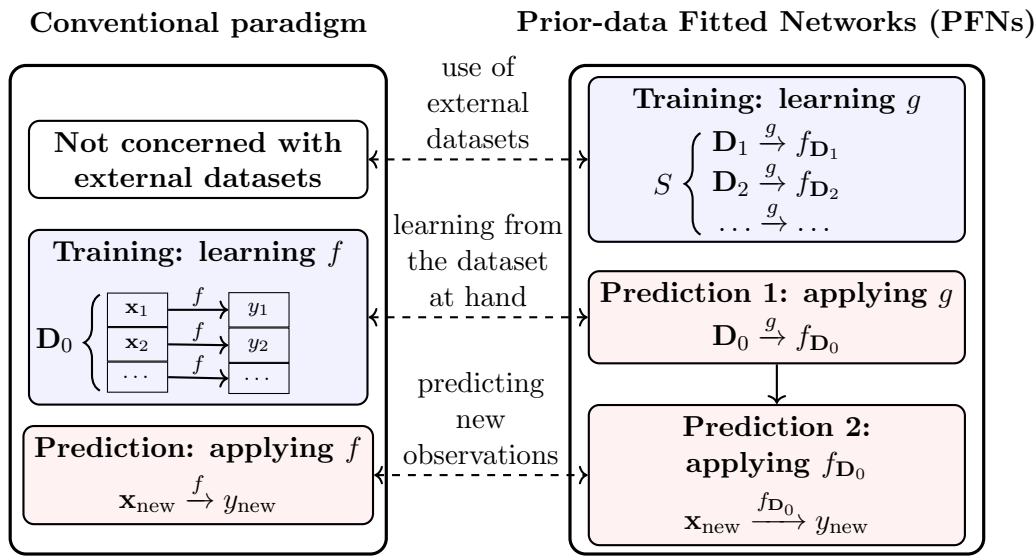
\begin{figure}[h]
\centering
\begin{tikzpicture}[
    font=\small,
    box/.style={draw, rounded corners=6pt, very thick, inner sep=6pt},
    subbox/.style={draw, rounded corners=4pt, thick, inner sep=4pt},
    arrow/.style={->, thick},
    dashedarrow/.style={<->, thick, dashed},
    every node/.style={align=center}
]

% --- LEFT: Conventional paradigm ---
\node[box, minimum width=5cm, minimum height=7cm] (conv) {};
\node[above=2mm of conv.north] {\textbf{Conventional paradigm}};

% Empty (left)
\node[subbox, minimum width=4.5cm] (conv-empty)
at ($(conv.center)+(0,1.9)$)
{\textbf{Not concerned with}\\\textbf{external datasets}};

% Training (left)
\node[subbox, fill=blue!5, minimum width=4.5cm] (conv-train) 
at ($(conv.center)+(0,-.4cm)$)
{
\textbf{Training: learning $f$}\\[1mm]
$\mathbf{x}_{0,1} \xrightarrow{f} \hat{y}_{0,1}$\\
$\mathbf{x}_{0,2} \xrightarrow{f} \hat{y}_{0,2}$\\
$\dots \xrightarrow{f} \dots$
};

\draw[
  decorate,
  decoration={brace, amplitude=6pt},
  thick
] 
($(conv-train.west)+(1.25cm,-1.1cm)$) -- ($(conv-train.west)+(1.25cm,0.35cm)$)
node[midway, xshift=-.6cm] {$\mathbf{D}_0$};

% Prediction (left)
\node[subbox, fill=red!5, minimum width=4.5cm] (conv-pred) 
at ($(conv.center)+(0,-2.55)$)
{\textbf{Prediction: applying $f$}\\[1mm]
$\mathbf{x}_{\text{new}} \xrightarrow{f} \hat{y}_{\text{new}}$
};

% --- RIGHT: PFN paradigm ---
\node[box, minimum width=7.5cm, minimum height=7cm, right=2.4cm of conv] (pfn) {};
\node[above=2mm of pfn.north] {\textbf{Prior-Data Fitted Networks (PFNs)}};

% Training (right, top)
\node[subbox, fill=blue!5, minimum width=7cm] (pfn-train)
at ($(pfn.center)+(0,1.9)$)
{\textbf{Training: learning $g$}\\
$\begin{aligned}
\begin{aligned}
\mathbf{D}_1 &\xrightarrow{g} f_{\mathbf{D}_1};\\
\mathbf{D}_2 &\xrightarrow{g} f_{\mathbf{D}_2};\\
\dots &\xrightarrow{g} \dots;
\end{aligned}
\quad
\begin{aligned}
(\mathbf{x}_{1,1},\dots) &\xrightarrow{\mathmakebox[.5cm]{f_{\mathbf{D}_1}}} (\hat{y}_{1,1},\dots)\\
(\mathbf{x}_{2,1},\dots) &\xrightarrow{\mathmakebox[.5cm]{f_{\mathbf{D}_2}}} (\hat{y}_{2,1},\dots)\\
(\dots) &\xrightarrow{\mathmakebox[.5cm]{\dots}} (\dots)
\end{aligned}
\end{aligned}$
};

% Prediction 1 (right, middle)
\node[subbox, fill=red!5, minimum width=7cm] (pfn-pred1)
at ($(pfn.center)+(0,-.4cm)$)
{\textbf{Prediction 1: applying $g$}\\[1mm]
$\mathbf{D}_0 \xrightarrow{g} f_{\mathbf{D}_0}$
};

% Prediction 2 (right, bottom)
\node[subbox, fill=red!5, minimum width=7cm] (pfn-pred2)
at ($(pfn.center)+(0,-2.55)$)
{\textbf{Prediction 2:} \textbf{applying $f_{\mathbf{D}_0}$}\\[1mm]
$\mathbf{x}_{\text{new}} \xrightarrow{f_{\mathbf{D}_0}} \hat{y}_{\text{new}}$
};

% --- Arrows ---
\draw[dashedarrow] (conv-empty.east) -- node[above] {use of\\external\\datasets} (pfn-train.west);
\draw[dashedarrow] (conv-train.east) -- node[above] {learning from\\the dataset\\at hand} (pfn-pred1.west);
\draw[dashedarrow] (conv-pred.east) -- node[above] {predicting\\new\\observations} (pfn-pred2.west);

\end{tikzpicture}
\caption{Comparison of learning paradigms.\\
$\mathbf{D}_1$, $\mathbf{D}_2, \dots$ is a collection of external datasets used to train the Prior-Data Fitted Network (PFN). $\mathbf{D}_0$ denotes the dataset from which one aims to develop a predictive algorithm  (e.g., a patient cohort). $\mathbf{x}_{j,i}$ denotes the vector of covariates of the patient $i$ and dataset $j$, and $\hat{y}_{j,i}$ denotes their predicted outcome. The learning phase aims to find a function that minimizes prediction error on held-out observations.}
\label{fig:paradigm}
\end{figure}

To better understand what makes this approach novel, we provide a conceptual description of PFNs and compare them with the conventional machine learning paradigm (as summarized in Figure~\ref{fig:paradigm}). In the latter, a model is trained on all observations from a single dataset with the goal of learning a function $f$ that maps an observation’s vector of covariates $\mathbf{x}$ to a predicted outcome $\hat{y}$. Learning $f$ amounts to finding a function that minimizes prediction error, by comparing the predicted outcome $f(\mathbf{x}) = \hat{y}$ with the actual outcome $y$ of held-out observations.
To better understand what makes this approach novel, we provide a conceptual description of PFNs and compare them with the conventional machine learning paradigm (as summarized in Figure~\ref{fig:paradigm}). In the latter, a model is trained on all observations from a single dataset with the goal of learning a function $f$ that maps an observation’s vector of covariates $\mathbf{x}$ to a predicted outcome $\hat{y}$. Learning $f$ amounts to finding a function that minimizes prediction error, by comparing the predicted outcome $f(\mathbf{x}) = \hat{y}$ with the actual outcome $y$ of held-out observations.

With PFNs, training is performed by their developers on a large \textit{set} of datasets to learn a function $g$ that maps an entire dataset $\mathbf{D}$ to a dataset-specific function $f_\mathbf{D}$ (the practical implications of this mechanism are described below). The function $f_\mathbf{D}$ maps an observation’s covariates $\mathbf{x}$ to a predicted outcome $\hat{y}$, just as in the conventional paradigm. Learning $g$ amounts to finding a function that minimizes prediction error, by comparing the predicted outcome $g(\mathbf{D})(\mathbf{x}) = f_\mathbf{D}(\mathbf{x}) = \hat{y}$ with the actual outcome $y$ of held-out observations.

To understand the implications of this paradigm shift, let us imagine using a PFN to develop a predictive model based on data from a patient cohort. The PFN has already been trained on numerous datasets by its developers. In epidemiological practice, the PFN's use of the cohort data therefore consists in applying the function $g$ it learned during training---the one that maps a dataset to a second function, which in turn maps an observation's covariates to a predicted outcome.

Thus, what would traditionally be considered a training phase on the cohort data is, in fact, already a prediction phase---the application of an already learned function. This form of learning during prediction time is called in-context learning (ICL)~\cite{garg_2022, bommasani_2022_foundationmodels, mueller2025mothernet}. ICL can be extremely fast, as parameters and hyperparameters are not modified during this step (no optimization or cross-validation is needed)~\cite{nagler_2023_statisticalfoundationspriordatafitted}. With TabPFN---a popular PFN---this phase takes about one second for datasets of $10,000$ observations and $10$ predictors~\cite{hollmann2025tabpfn}. The conventional prediction of an outcome for a new observation based on its covariates then follows as the next prediction step, corresponding to the PFN’s application of its second function ($f_\mathbf{D}$).

Naturally, the performance of a PFN depends on its training data, which should encompass scenarios similar to those the model will encounter during its use in clinical research, for example~\cite{nagler_2023_statisticalfoundationspriordatafitted, hollmann2025tabpfn}.
For TabPFN, the training data consist of millions of synthetic datasets, each generated from a randomly chosen causal structure and a randomly chosen set of functions, together formalized as a structural causal model ~\cite{hollmann2025tabpfn,pearl_2009_causality}. Compared to previous PFN versions, TabPFN allows for larger datasets and accepts categorical variables and missing values~\cite{muller2022transformers, hollmann_2023_tabpfn}. 
For prediction tasks involving tabular data, TabPFN frequently outperforms state-of-the-art algorithms, as measured by metrics such as area under the receiver operating characteristic curve (ROC AUC) for classification and negative root mean squared error (RMSE) for regression, for datasets containing up to $10,000$ observations and $500$ features~\cite{hollmann2025tabpfn, zhang2025tabpfnmodelruleall}.

The predictive performance of TabPFN could be leveraged in causal inference settings, where flexible supervised learning methods are employed through procedures such as inverse probability of treatment weighting (IPTW) and g-computation (model-based standardization) to estimate marginal (population-level) treatment effects~\cite{rosenbaum_1983_centralroleofpropensityscore, robins_1986_newapproachtocausalinference, pearl_1995_causaldiagrams, snowden2010g-computation}. In this field, machine learning is notably used to automate part of the statistical modeling process by introducing parametric constraints to make estimation possible~\cite{Baiardi2024_valueaddedofmachinelearningtocausalinference}.

However, machine learning algorithms are optimized for prediction, which involves accepting a certain level of bias to reduce the variance of their predictions. This bias carries over to the causal inference stage, motivating the development of bias-correction methods such as ``double/debiased" machine learning or doubly robust approaches~\cite{Chernozhukov_2018_doubleML_for_trt_and_structural_parameters, naimi_2023_challenges_valide_causal_effect_ML, bang_2005_doublyrobustestimation, vanderLaan_2006_tmle, Funk_2011}.

A specific bias related to g-computation arises when the machine learning algorithm downweights the treatment variable in predicting the outcome, causing the estimated treatment effect to shrink toward zero~\cite{Kunzel2019_metalearnersforhtesusingml}. A common solution is to train a separate model for each treatment group, corresponding to the so-called T-learner approach.

Computation time is another limitation of these approaches. In addition to hyperparameter tuning, frequently involving cross-validation, confidence intervals are often constructed using procedures such as bootstrap. As a result, repeating hyperparameter tuning at each bootstrap iteration can lead to considerable computational costs. While simpler approaches based on standard generalized linear models without tuning may alleviate this burden, they rely on correct specification of functional forms and interactions, which can be challenging in high-dimensional settings or when prior knowledge is limited. In this context, PFNs may offer an alternative, by reducing computational demand while retaining the ability to capture complex relationships.

Beyond the statistical modeling step, some PFNs aim to automate causal modeling itself by providing an estimate of the marginal treatment effect directly from a dataset, without the need for methods such as g-computation or IPTW~\cite{balazadeh2025causalpfn,robertson2025dopfn}. These PFNs also provide credible intervals for the treatment effect as a by-product of their computations, eliminating the computation time associated with bootstrapping.

In this study, we evaluated the applicability and performance of PFNs to estimate the average treatment effect (ATE) using real-world simulation scenarios that satisfy causal identifiability assumptions (positivity, conditional exchangeability, and consistency). We assessed TabPFN through IPTW and g-computation, and CausalPFN, which directly provides an estimate of the ATE. As comparators, we used XGBoost---a widely used gradient boosting tree algorithm---without hyperparameter tuning (to limit computational burden), and logistic regression without interaction terms (reflecting common practice), both used for IPTW and g-computation~\cite{R_xgboost}.

\section*{Methods}\label{methods}

\subsection*{Causal setting}

\begin{figure}[h]
\centering
\includegraphics[width=.4\textwidth]{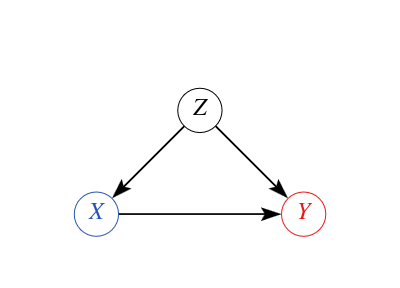}
\caption{Causal graph representing the causal inference task, where $X$ denotes treatment, $Y$ denotes outcome, and $Z$ is a set of common causes of both $X$ and $Y$.}\label{dag}
\end{figure}

We focused on estimating the ATE of a non–time-varying binary treatment ($X$) on a binary outcome ($Y$), in the presence of categorical common causes ($Z$) of both treatment and outcome, i.e., confounding factors, as depicted in Figure~\ref{dag}.

The ATE was considered on the absolute difference scale; that is, the ATE was the difference between the expected counterfactual outcome under treatment ($Y^{X=1}$) and the expected counterfactual outcome without treatment ($Y^{X=0}$)~\cite{whatif}:
$$\text{ATE} = E[Y^{X=1}] - E[Y^{X=0}]$$

\subsection*{Data and Simulation}\label{simulation}

We designed two simulation scenarios, each based on a real-world medical dataset from the \texttt{medicaldata} and \texttt{survival} R packages~\cite{medicaldata,survival_package}. To this end, we first prepared the datasets and then used their empirical distributions to derive the simulation parameters, as described below.

The data preparation aimed to ensure realism while satisfying the assumptions of positivity, consistency, and conditional exchangeability. For each scenario, we defined an outcome variable ($Y$), a treatment variable ($X$), and a set of confounders. These confounders were first coded as binary variables: continuous variables were dichotomized at their median, and categorical variables with more than two levels were collapsed into two groups with the objective of preserving balanced proportions. We then defined $Z$ as a categorical variable formed by the joint values of the binary confounders, representing all possible combinations of these values. $Z$ being a categorical variable, we were able to verify the positivity assumption. This was done by ensuring that each level of $Z$ in the dataset included at least one individual receiving each level of the treatment $X$. Levels of $Z$ that did not meet this criterion were excluded from the scenarios. The resulting datasets are described in the Results section.

Each simulation scenario was defined by the empirical distributions $${P(Z),P(X|z),P(Y|x,z)}$$ (provided in Supplementary material~A), calculated directly from the corresponding dataset after the data preparation described above. For example, $P(X=1|z)$ corresponded to the proportion of patients who received treatment ($X=1$) among those with a confounder profile $z$ in the dataset. To simulate a single observation in a given scenario, we first drew a value $z$ for the variable $Z$ from the multinomial distribution ${P(Z)}$. Next, we drew a value $x$ for the variable $X$ from the Bernoulli distribution ${P(X|z)}$, with the probability of success depending on $z$. Finally, we drew a value $y$ for the variable $Y$ from the Bernoulli distribution ${P(Y|x,z)}$, with the probability of success depending on both $x$ and $z$. To simulate a sample of size $n$, we repeated this procedure $n$ times, drawing from the same distributions each time. For each scenario, ${1,000}$ datasets were simulated for each sample size (${n=200}$, ${n=500}$, and ${n=1,000}$).

With this simulation procedure, the causal identification conditions of positivity, conditional exchangeability, and consistency hold~\cite{whatif}. More precisely, ${0<P(X=1|z)<1}$ holds for each $z$ (positivity), ${Y^x \indep X|Z}$ (conditional exchangeability) holds as treatment assignment is conditionally randomized given $Z$~\cite{rosenbaum_1983_centralroleofpropensityscore, whatif}, and the treatment levels have a unique version (consistency). Finally, the outcome of a patient is not influenced by the treatment of other patients (no interference).

\subsection*{Estimators}

We compared several estimators of the ATE, each combining a causal inference strategy and a statistical model. Causal inference strategies were g-computation and IPTW (implemented as described below). The three statistical models used were:
\begin{itemize}
    \item TabPFN
    \item A logistic regression model without interaction terms, including one parameter for each binary confounder, one for the treatment, and an intercept (we emphasize that this model is misspecified, as the simulation method preserves natural interactions in the data)
    \item Extreme Gradient Boosting (XGBoost), using default hyperparameter settings (i.e., without hyperparameter tuning)
\end{itemize}

CausalPFN was implemented without any external causal inference strategy, as it automates this step and directly provides an estimate of the ATE. As a reference, we also assessed the performance of the difference in outcome means between the two treatment groups without any adjustment method, which captures the crude association between treatment and outcome.

The 95\% confidence intervals (CI) were constructed using the bootstrap percentile method with 599 iterations for all approaches, except for CausalPFN, which natively provides its own 95\% credible intervals~\cite{balazadeh2025causalpfn}.

\subsubsection*{G-computation}\label{g_comp}

G-computation, also known as back-door adjustment, is a method for estimating causal effects by standardizing the expected outcome across levels of some covariates that satisfy the assumption of conditional exchangeability~\cite{whatif, pearl_2009_causality}. We implemented a single-model g-computation to estimate the ATE in three steps~\cite{snowden2010g-computation}:
\begin{enumerate}
    \item Fit a model where the outcome is regressed on treatment and confounding variables to provide \({\hat{E}[Y|x,z]}\) for all $x$ and $z$.
    \item For each individual \(i\), predict the outcomes under both treatment conditions by setting \({X=0}\) and \({X=1}\), while holding the confounders \(Z_i\) at their observed values \(z_i\). That is, compute \({\hat{E}[Y|X=0,z_i]}\) and \({\hat{E}[Y|X=1,z_i]}\) for each individual.
    \item Estimate the ATE by averaging individual-level contrasts (i.e., conditional average treatment effects):
    $$\widehat{\text{ATE}} = \frac{1}{n}\sum_{i=1}^n \Bigl(\hat{E}[Y|X=1, z_i] - \hat{E}[Y|X=0, z_i]\Bigr)$$
\end{enumerate}

We also implemented a two-model outcome specification g-computation, known as T-learner in the machine learning community, allowing separate outcome models for the treated and control groups (this amounts to introduce interaction terms between $X$ and $Z$)~\cite{Salditt2024_hatewithmetalearners}:
\begin{enumerate}
\item Separate the sample according to treatment status ($X=0$ or $X=1$). For each subsample, fit a model where the outcome is regressed on confounding variables to provide ${\hat{E}[Y|X=0,z]}$ and ${\hat{E}[Y|X=1,z]}$, respectively. 
\item For each individual $i$, predict outcomes using both models, based on their observed covariates $z_i$.
\item Estimate the ATE as in the single-model approach.

\end{enumerate}

\subsubsection*{Inverse probability of treatment weighting}\label{iptw}

IPTW differs from outcome modeling by focusing on the treatment assignment mechanism. It estimates causal effects by weighting each individual by the inverse probability of receiving their observed treatment, conditional on their covariates $Z$. IPTW was implemented in three steps~\cite{Kostouraki_2024}:
\begin{enumerate}
    \item For each individual $i$, compute \({\hat{e}_i = \hat{P}(X = 1|z_i)}\), the estimated probability of receiving the treatment given $z_i$.
    \item Compute a weight $w_i$ for each individual $i$ as follows:
    $$
    w_i = \begin{cases}
    \frac{1}{\hat{e}_i} & \text{if } X_i = 1 \\
    \frac{1}{1 - \hat{e}_i} & \text{if } X_i = 0
    \end{cases}
    $$
    \item The ATE was estimated using the Hájek estimator, by comparing the weighted means of the outcome between treatment groups:
    $$\widehat{\text{ATE}} = \frac{\sum_{i=1}^n w_iX_iY_i}{\sum_{i=1}^n w_iX_i}- \frac{\sum_{i=1}^n w_i(1-X_i)Y_i}{\sum_{i=1}^n w_i(1-X_i)}
    $$
\end{enumerate}

\subsection*{Performance metrics}\label{performance_criteria}

For each scenario, we computed the true ATE using the back-door adjustment formula for each level of $X$ (using the distributions $P(Z)$, $P(X|z)$, and $P(Y|x, z)$ that defined the simulation procedure)~\cite{pearl_1995_causaldiagrams}:
\begin{align*}
\text{ATE}&=E[Y^{X=1}] - E[Y^{X=0}]\\
&=\sum_z P(Y=1|X=1, z)P(z) - \sum_z P(Y=1|X=0, z)P(z)
\end{align*}
We evaluated the performance of each approach using the following metrics (formally defined in Supplementary material~B): 95\% confidence interval coverage (or 95\% credible interval coverage, in the case of CausalPFN), mean 95\% CI width, bias-eliminated 95\% CI coverage (i.e., coverage after centering estimates at the true value)~\cite{Morris2019_simulation_statistical_methods}, mean error (i.e., the difference between the true and estimated ATEs), mean squared error (MSE), and average computation time required to obtain a point estimate along with its confidence interval.

\subsection*{Software and hardware}

TabPFN was implemented using the \texttt{tabpfn} Python package, version~2.1.0~\cite{hollmann2025tabpfn}.
CausalPFN was implemented using \texttt{causalpfn}, version~0.1.4~\cite{balazadeh2025causalpfn}.
XGBoost was implemented using the \texttt{xgboost} package in R~\cite{R_xgboost}.
We used R version 4.4.2~\cite{R} for all analyses, except for running TabPFN and CausalPFN, which were executed with Python version 3.10.18~\cite{hollmann2025tabpfn, python}.

All computations were performed on a system equipped with an Apple M4 Pro processor. Computation with TabPFN took considerably longer than anticipated, prompting the use of a graphics processing unit (GPU) Nvidia A100 Tensor. The reported computation time for TabPFN, however, refers to a single simulation run on an Apple M4 Pro processor. Our code and simulated data are available at \href{https://github.com/franciscomourao/PFNs_Causal_Inference}{https://github.com/franciscomourao/PFNs\_Causal\_Inference}.

\section*{Results}\label{results}
\subsection*{Real-World Datasets}\label{datasets}

\subsubsection*{Indomethacin}
Detailed characteristics of the datasets used to build the simulation scenarios are provided in Supplementary material~C. The Indomethacin dataset is open access, available in R through the \texttt{medicaldata} package. It presents the results of a randomized, placebo-controlled trial of indomethacin 100~mg administered rectally versus placebo to prevent post-endoscopic retrograde cholangiopancreatography (ERCP) pancreatitis and involved 602 patients~\cite{elmunzer_2012_rctindomethacin}.

We focused on the effect of pancreatic sphincterotomy ($X$, non-randomized) on post-ERCP pancreatitis ($Y$). We selected age, gender, study site (restricted to the two sites with the largest patient enrollment), and sphincter of Oddi dysfunction as components of $Z$. Covariate strata without representation of both treatment levels were removed, yielding $14$ unique covariate $Z$ combinations. The final sample size was ${n=570}$, with 331 patients receiving a pancreatic sphincterotomy, and the incidence of post-ERCP pancreatitis was $13.5\%$. The true ATE for this scenario was $-0.062$.

\subsubsection*{Rotterdam}
The Rotterdam dataset is open-access, available in R through the \texttt{survival} package. It included ${2,982}$ patients with primary breast cancers from the Rotterdam tumor bank~\cite{Royston_2013}.

We focused on the effect of adjuvant chemotherapy ($X$) on the composite outcome of death or recurrence at 5 years after primary surgery. We selected year of surgery, age, presence of positive lymph nodes, progesterone receptors, estrogen receptors, menopausal status, tumor size, and differentiation grade as components of $Z$. We excluded patients who were lost to follow-up before 5 years (approximately 4\% of the cohort). Additionally, we excluded patients with covariate patterns that did not have both treatment levels, resulting in $98$ unique covariate $Z$ combinations. The final sample size was ${n=2,260}$, 23\% of patients received chemotherapy, and the incidence of the outcome was 46\%. The true ATE for this scenario was $-0.101$.

\subsection*{Confidence Interval Metrics}

\begin{figure}[ht]%
\centering
  \includegraphics[width=0.9\textwidth]{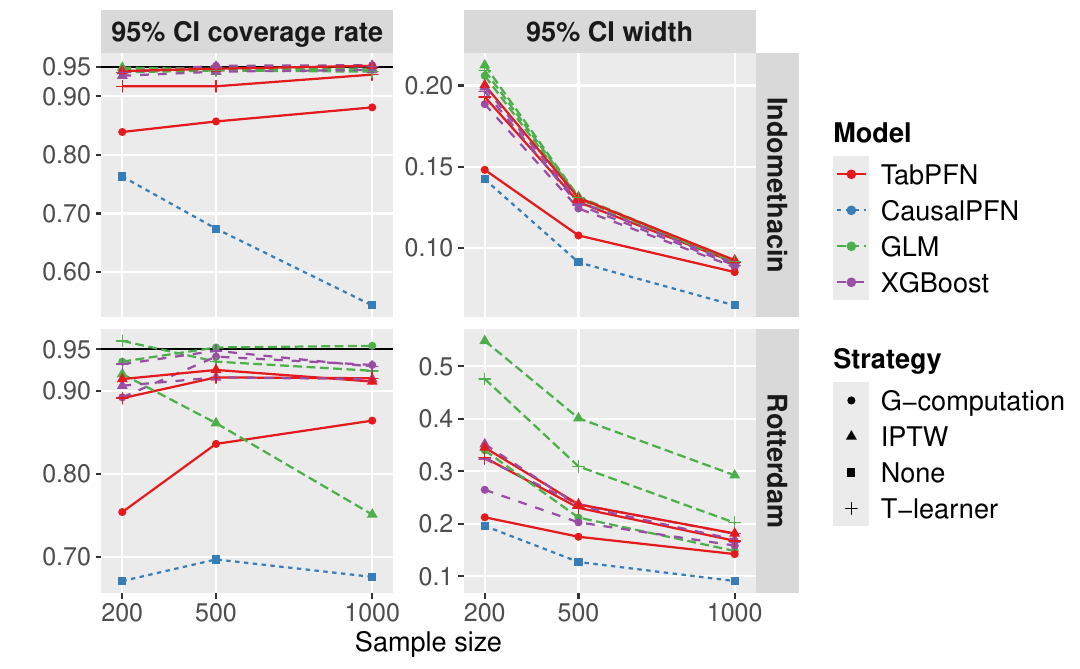}
    \captionof{figure}{Confidence interval metrics for the two simulation scenarios. CI: confidence interval. GLM: logistic regression without interaction terms. IPTW: inverse probability of treatment weighting. G-computation corresponds to g-computation with a single model. T-learner corresponds to g-computation with one model per treatment group. The performance of the ``crude association" approach is not shown in this figure.}\label{coverage_plots}
\end{figure}

Exhaustive simulation results are provided in Table~\ref{table_indomethacin} and Table~\ref{table_rotterdam}; Figure~\ref{coverage_plots} shows the results for confidence interval metrics.

% Tableau résultats Indomethacin
\begin{table}[!h]
\centering
\caption{Simulation results by method for the Indomethacin scenario (1,000 simulations, 599 bootstrap iterations).}
\label{table_indomethacin}
\centering
\resizebox{\ifdim\width>\linewidth\linewidth\else\width\fi}{!}{
\fontsize{7}{9}\selectfont
\begin{tabular}[t]{l>{\centering\arraybackslash}p{1.5cm}>{\centering\arraybackslash}p{1.5cm}>{\centering\arraybackslash}p{1.5cm}>{\centering\arraybackslash}p{1.5cm}>{\centering\arraybackslash}p{1.5cm}>{\centering\arraybackslash}p{1.5cm}>{}p{1.5cm}}
\toprule
Method & 95\% CI Coverage & Mean 95\% width & 95\% CI Coverage Bias Eliminated & Mean Error ($10^3$) & Mean Squared Error ($10^3$) & Mean Computation Time (s)\\
\midrule
\addlinespace[0.3em]
\hline
\multicolumn{7}{l}{\cellcolor{lightgray}{\textbf{Sample size n = 200}}}\\
\hline
\addlinespace[0.5em]
\multicolumn{7}{l}{\textbf{Crude association}}\\
\hspace{1em}\hspace{1em}$\mathbb{E}(Y|X=1) - \mathbb{E}(Y|X=0)$ & 76.8 & 19 & 93.8 & 58.91 & 5.92 & \textcolor{orange}{\textbf{2}}\\
\addlinespace[0.5em]
\multicolumn{7}{l}{\textbf{G-computation}}\\
\hspace{1em}\hspace{1em}GLM & \textcolor{orange}{\textbf{94.5}} & 20.6 & 94.6 & 5.49 & 2.84 & \textcolor{blue}{\textbf{3}}\\
\hspace{1em}\hspace{1em}XGBoost & 94.1 & \textcolor{blue}{\textbf{18.8}} & \textcolor{red}{\textbf{95.1}}\textsuperscript{*} & 9.67 & \textcolor{red}{\textbf{2.43}} & 13\\
\hspace{1em}\hspace{1em}TabPFN & 83.9 & \textcolor{orange}{\textbf{14.8}} & \textcolor{red}{\textbf{94.9}}\textsuperscript{*} & 43.29 & \textcolor{blue}{\textbf{2.8}} & 1380\\
\addlinespace[0.5em]
\multicolumn{7}{l}{\textbf{Two-model G-computation}}\\
\hspace{1em}\hspace{1em}GLM & 94.1 & 20.9 & 94.2 & 6.35 & 2.91 & 10\\
\hspace{1em}\hspace{1em}XGBoost & 94.1 & 19.6 & \textcolor{red}{\textbf{94.9}}\textsuperscript{*} & 9.96 & \textcolor{orange}{\textbf{2.68}} & 42\\
\hspace{1em}\hspace{1em}TabPFN & 91.7 & 19.3 & 94.1 & 31.75 & 3.58 & 1208\\
\addlinespace[0.5em]
\multicolumn{7}{l}{\textbf{IPTW}}\\
\hspace{1em}\hspace{1em}GLM & \textcolor{red}{\textbf{94.9}} & 21.2 & 94.7 & \textcolor{blue}{\textbf{4.78}} & 2.91 & 4\\
\hspace{1em}\hspace{1em}XGBoost & 93.5 & 19.9 & 93.6 & \textcolor{orange}{\textbf{4.13}} & 2.86 & 13\\
\hspace{1em}\hspace{1em}TabPFN & \textcolor{blue}{\textbf{94.3}} & 20 & 94.3 & \textcolor{red}{\textbf{3.88}} & 2.81 & 785\\
\addlinespace[0.5em]
\multicolumn{7}{l}{\textbf{CausalPFN}}\\
\hspace{1em}\hspace{1em}CausalPFN & 76.3 & \textcolor{red}{\textbf{14.2}} & 84.7 & 29.53 & 3.39 & \textcolor{red}{\textbf{1}}\\
\addlinespace[0.3em]
\hline
\multicolumn{7}{l}{\cellcolor{lightgray}{\textbf{Sample size n = 500}}}\\
\hline
\addlinespace[0.5em]
\multicolumn{7}{l}{\textbf{Crude association}}\\
\hspace{1em}\hspace{1em}$\mathbb{E}(Y|X=1) - \mathbb{E}(Y|X=0)$ & 47.8 & \textcolor{blue}{\textbf{12}} & 94.3 & 61.22 & 4.68 & \textcolor{red}{\textbf{2}}\\
\addlinespace[0.5em]
\multicolumn{7}{l}{\textbf{G-computation}}\\
\hspace{1em}\hspace{1em}GLM & \textcolor{blue}{\textbf{94.7}} & 13 & 95.2 & 7.59 & 1.15 & \textcolor{orange}{\textbf{3}}\\
\hspace{1em}\hspace{1em}XGBoost & \textcolor{orange}{\textbf{95.2}} & 12.4 & \textcolor{red}{\textbf{95}}\textsuperscript{*} & 5.05 & \textcolor{red}{\textbf{1.02}} & 14\\
\hspace{1em}\hspace{1em}TabPFN & 85.7 & \textcolor{orange}{\textbf{10.8}} & 95.2 & 35.14 & 1.88 & 4977\\
\addlinespace[0.5em]
\multicolumn{7}{l}{\textbf{Two-model G-computation}}\\
\hspace{1em}\hspace{1em}GLM & 94.4 & 13.1 & \textcolor{red}{\textbf{95}}\textsuperscript{*} & 8.02 & 1.17 & 4\\
\hspace{1em}\hspace{1em}XGBoost & \textcolor{red}{\textbf{95}} & 12.6 & 94.6 & 6.59 & \textcolor{orange}{\textbf{1.09}} & 21\\
\hspace{1em}\hspace{1em}TabPFN & 91.7 & 12.8 & \textcolor{blue}{\textbf{94.9}} & 20.25 & 1.52 & 3151\\
\addlinespace[0.5em]
\multicolumn{7}{l}{\textbf{IPTW}}\\
\hspace{1em}\hspace{1em}GLM & 94.6 & 13.1 & 94.6 & 5.83 & 1.15 & 4\\
\hspace{1em}\hspace{1em}XGBoost & 94.2 & 12.9 & 94.5 & \textcolor{blue}{\textbf{2.52}} & \textcolor{blue}{\textbf{1.12}} & 13\\
\hspace{1em}\hspace{1em}TabPFN & \textcolor{blue}{\textbf{94.7}} & 13.1 & 94.6 & \textcolor{orange}{\textbf{-0.36}} & 1.13 & 2380\\
\addlinespace[0.5em]
\multicolumn{7}{l}{\textbf{CausalPFN}}\\
\hspace{1em}\hspace{1em}CausalPFN & 67.4 & \textcolor{red}{\textbf{9.1}} & 84.3 & 30.89 & 1.98 & \textcolor{blue}{\textbf{3}}\\
\addlinespace[0.3em]
\hline
\multicolumn{7}{l}{\cellcolor{lightgray}{\textbf{Sample size n = 1000}}}\\
\hline
\addlinespace[0.5em]
\multicolumn{7}{l}{\textbf{Crude association}}\\
\hspace{1em}\hspace{1em}$\mathbb{E}(Y|X=1) - \mathbb{E}(Y|X=0)$ & 19 & \textcolor{orange}{\textbf{8.5}} & \textcolor{red}{\textbf{95}}\textsuperscript{*} & 59.91 & 4.03 & \textcolor{red}{\textbf{2}}\\
\addlinespace[0.5em]
\multicolumn{7}{l}{\textbf{G-computation}}\\
\hspace{1em}\hspace{1em}GLM & 94.5 & 9.1 & 94.5 & 6.84 & 0.54 & \textcolor{orange}{\textbf{4}}\\
\hspace{1em}\hspace{1em}XGBoost & \textcolor{blue}{\textbf{95.3}} & 8.8 & \textcolor{red}{\textbf{95}}\textsuperscript{*} & \textcolor{orange}{\textbf{2.29}} & \textcolor{red}{\textbf{0.47}} & 18\\
\hspace{1em}\hspace{1em}TabPFN & 88.1 & \textcolor{blue}{\textbf{8.5}} & 95.9 & 25.74 & 1.13 & 10879\\
\addlinespace[0.5em]
\multicolumn{7}{l}{\textbf{Two-model G-computation}}\\
\hspace{1em}\hspace{1em}GLM & 94.2 & 9.2 & \textcolor{blue}{\textbf{95.2}} & 7.37 & 0.55 & 5\\
\hspace{1em}\hspace{1em}XGBoost & \textcolor{red}{\textbf{95.1}} & 8.9 & 94.7 & 3.66 & \textcolor{orange}{\textbf{0.49}} & 23\\
\hspace{1em}\hspace{1em}TabPFN & 93.7 & 9.1 & 95.7 & 11.56 & 0.63 & 5970\\
\addlinespace[0.5em]
\multicolumn{7}{l}{\textbf{IPTW}}\\
\hspace{1em}\hspace{1em}GLM & \textcolor{blue}{\textbf{94.7}} & 9.2 & 95.3 & 5.22 & 0.53 & \textcolor{blue}{\textbf{5}}\\
\hspace{1em}\hspace{1em}XGBoost & 94.5 & 9 & 94.7 & \textcolor{red}{\textbf{0.56}} & \textcolor{blue}{\textbf{0.49}} & 16\\
\hspace{1em}\hspace{1em}TabPFN & \textcolor{orange}{\textbf{95.2}} & 9.2 & 95.4 & \textcolor{blue}{\textbf{-3.2}} & 0.53 & 5885\\
\addlinespace[0.5em]
\multicolumn{7}{l}{\textbf{CausalPFN}}\\
\hspace{1em}\hspace{1em}CausalPFN & 54.3 & \textcolor{red}{\textbf{6.5}} & 84.2 & 30.73 & 1.45 & 10\\
\bottomrule
\end{tabular}}
\begin{tablenotes}%
\tiny
\item CI: confidence interval; GLM: logistic regression without interaction terms; IPTW: inverse probability of treatment weighting.
\item Color legend: \textbf{\textcolor{red}{Red}} = best, \textbf{\textcolor{orange}{Orange}} = second best, \textbf{\textcolor{blue}{Blue}} = third best.
\item *Tied for best result. 
\end{tablenotes}
\end{table}

% Tableau résultats Rotterdam
\begin{table}[!h]
\centering
\caption{Simulation results by method for the Rotterdam scenario (1,000 simulations, 599 bootstrap iterations).}
\label{table_rotterdam}
\centering
\resizebox{\ifdim\width>\linewidth\linewidth\else\width\fi}{!}{
\fontsize{7}{9}\selectfont
\begin{tabular}[t]{l>{\centering\arraybackslash}p{1.5cm}>{\centering\arraybackslash}p{1.5cm}>{\centering\arraybackslash}p{1.5cm}>{\centering\arraybackslash}p{1.5cm}>{\centering\arraybackslash}p{1.5cm}>{\centering\arraybackslash}p{1.5cm}>{}p{1.5cm}}
\toprule
Method & 95\% CI Coverage & Mean 95\% width & 95\% CI Coverage Bias Eliminated & Mean Error ($10^3$) & Mean Squared Error ($10^3$) & Mean Computation Time (s)\\
\midrule
\addlinespace[0.3em]
\hline
\multicolumn{7}{l}{\cellcolor{lightgray}{\textbf{Sample size n = 200}}}\\
\hline
\addlinespace[0.5em]
\multicolumn{7}{l}{\textbf{Crude association}}\\
\hspace{1em}\hspace{1em}$\mathbb{E}(Y|X=1) - \mathbb{E}(Y|X=0)$ & 74.8 & 32.7 & 93.5 & 108.39 & 19.2 & \textcolor{orange}{\textbf{2}}\\
\addlinespace[0.5em]
\multicolumn{7}{l}{\textbf{G-computation}}\\
\hspace{1em}\hspace{1em}GLM & \textcolor{orange}{\textbf{93.5}} & 34 & 93.5 & \textcolor{red}{\textbf{-1.13}} & \textcolor{blue}{\textbf{7.8}} & 6\\
\hspace{1em}\hspace{1em}XGBoost & 89.2 & \textcolor{blue}{\textbf{26.5}} & \textcolor{red}{\textbf{95}} & 35.54 & \textcolor{red}{\textbf{6.93}} & 25\\
\hspace{1em}\hspace{1em}TabPFN & 75.4 & \textcolor{orange}{\textbf{21.2}} & 93.5 & 74.5 & \textcolor{orange}{\textbf{7.59}} & 2665\\
\addlinespace[0.5em]
\multicolumn{7}{l}{\textbf{Two-model G-computation}}\\
\hspace{1em}\hspace{1em}GLM & \textcolor{red}{\textbf{96}} & 47.6 & \textcolor{blue}{\textbf{96}} & \textcolor{orange}{\textbf{-6.36}} & 19.46 & \textcolor{blue}{\textbf{6}}\\
\hspace{1em}\hspace{1em}XGBoost & \textcolor{blue}{\textbf{93.2}} & 32.4 & \textcolor{orange}{\textbf{94.7}} & -28.66 & 8.89 & 23\\
\hspace{1em}\hspace{1em}TabPFN & 89.1 & 32.6 & 92.2 & 49.49 & 11.46 & 2324\\
\addlinespace[0.5em]
\multicolumn{7}{l}{\textbf{IPTW}}\\
\hspace{1em}\hspace{1em}GLM & 92 & 54.8 & 92.6 & -47.42 & 33.13 & 8\\
\hspace{1em}\hspace{1em}XGBoost & 90.6 & 35.2 & 93 & 33.58 & 12.81 & 24\\
\hspace{1em}\hspace{1em}TabPFN & 91.4 & 34.5 & 92 & \textcolor{blue}{\textbf{10.94}} & 12.09 & 1471\\
\addlinespace[0.5em]
\multicolumn{7}{l}{\textbf{CausalPFN}}\\
\hspace{1em}\hspace{1em}CausalPFN & 67.1 & \textcolor{red}{\textbf{19.6}} & 69.2 & 30.37 & 10.34 & \textcolor{red}{\textbf{1}}\\
\addlinespace[0.3em]
\hline
\multicolumn{7}{l}{\cellcolor{lightgray}{\textbf{Sample size n = 500}}}\\
\hline
\addlinespace[0.5em]
\multicolumn{7}{l}{\textbf{Crude association}}\\
\hspace{1em}\hspace{1em}$\mathbb{E}(Y|X=1) - \mathbb{E}(Y|X=0)$ & 50.8 & 20.7 & \textcolor{red}{\textbf{95}} & 104.4 & 13.67 & \textcolor{red}{\textbf{2}}\\
\addlinespace[0.5em]
\multicolumn{7}{l}{\textbf{G-computation}}\\
\hspace{1em}\hspace{1em}GLM & \textcolor{red}{\textbf{95.2}}\textsuperscript{*} & 21.2 & \textcolor{orange}{\textbf{95.2}} & \textcolor{red}{\textbf{-3.92}} & \textcolor{red}{\textbf{3}} & \textcolor{blue}{\textbf{4}}\\
\hspace{1em}\hspace{1em}XGBoost & \textcolor{blue}{\textbf{94.1}} & \textcolor{blue}{\textbf{20.3}} & \textcolor{blue}{\textbf{95.3}} & 14.03 & \textcolor{orange}{\textbf{3.29}} & 17\\
\hspace{1em}\hspace{1em}TabPFN & 83.6 & \textcolor{orange}{\textbf{17.5}} & 93.4 & 51.5 & 4.47 & 6838\\
\addlinespace[0.5em]
\multicolumn{7}{l}{\textbf{Two-model G-computation}}\\
\hspace{1em}\hspace{1em}GLM & 93.5 & 30.9 & 93.8 & \textcolor{orange}{\textbf{3.96}} & 6.19 & 6\\
\hspace{1em}\hspace{1em}XGBoost & \textcolor{red}{\textbf{94.8}}\textsuperscript{*} & 23.7 & 95.4 & \textcolor{blue}{\textbf{-10.96}} & 4.46 & 26\\
\hspace{1em}\hspace{1em}TabPFN & 91.6 & 23 & 92.9 & 20.7 & 4.3 & 4598\\
\addlinespace[0.5em]
\multicolumn{7}{l}{\textbf{IPTW}}\\
\hspace{1em}\hspace{1em}GLM & 86.1 & 40.1 & 93.2 & -79.79 & 20.33 & 5\\
\hspace{1em}\hspace{1em}XGBoost & 91.6 & 23.5 & 92.7 & 12.96 & 5.22 & 16\\
\hspace{1em}\hspace{1em}TabPFN & 92.5 & 23.7 & 91.4 & -13.51 & 5.78 & 3644\\
\addlinespace[0.5em]
\multicolumn{7}{l}{\textbf{CausalPFN}}\\
\hspace{1em}\hspace{1em}CausalPFN & 69.7 & \textcolor{red}{\textbf{12.7}} & 72.6 & 19.25 & \textcolor{blue}{\textbf{4.18}} & \textcolor{orange}{\textbf{3}}\\
\addlinespace[0.3em]
\hline
\multicolumn{7}{l}{\cellcolor{lightgray}{\textbf{Sample size n = 1000}}}\\
\hline
\addlinespace[0.5em]
\multicolumn{7}{l}{\textbf{Crude association}}\\
\hspace{1em}\hspace{1em}$\mathbb{E}(Y|X=1) - \mathbb{E}(Y|X=0)$ & 19.7 & \textcolor{blue}{\textbf{14.6}} & \textcolor{orange}{\textbf{94.6}} & 105.19 & 12.46 & \textcolor{red}{\textbf{2}}\\
\addlinespace[0.5em]
\multicolumn{7}{l}{\textbf{G-computation}}\\
\hspace{1em}\hspace{1em}GLM & \textcolor{red}{\textbf{95.4}} & 14.8 & \textcolor{red}{\textbf{95.2}} & \textcolor{orange}{\textbf{-4.37}} & \textcolor{red}{\textbf{1.41}} & \textcolor{orange}{\textbf{5}}\\
\hspace{1em}\hspace{1em}XGBoost & \textcolor{orange}{\textbf{93.1}} & 15.9 & \textcolor{blue}{\textbf{94.5}} & 14.05 & \textcolor{blue}{\textbf{2.14}} & 22\\
\hspace{1em}\hspace{1em}TabPFN & 86.4 & \textcolor{orange}{\textbf{14.2}} & 91.7 & 35.68 & 2.9 & 16504\\
\addlinespace[0.5em]
\multicolumn{7}{l}{\textbf{Two-model G-computation}}\\
\hspace{1em}\hspace{1em}GLM & 92.4 & 20.2 & 92.6 & \textcolor{red}{\textbf{1.66}} & 2.81 & 6\\
\hspace{1em}\hspace{1em}XGBoost & \textcolor{blue}{\textbf{92.9}} & 18.1 & 93.7 & -12.15 & 2.65 & 29\\
\hspace{1em}\hspace{1em}TabPFN & 91.5 & 16.8 & 92.3 & 11.45 & 2.17 & 9320\\
\addlinespace[0.5em]
\multicolumn{7}{l}{\textbf{IPTW}}\\
\hspace{1em}\hspace{1em}GLM & 75.1 & 29.2 & 91.4 & -81.68 & 13.33 & \textcolor{blue}{\textbf{6}}\\
\hspace{1em}\hspace{1em}XGBoost & 91.4 & 17 & 93 & \textcolor{blue}{\textbf{9.25}} & 2.68 & 20\\
\hspace{1em}\hspace{1em}TabPFN & 91.1 & 18.2 & 91.2 & -15.64 & 3.19 & 7863\\
\addlinespace[0.5em]
\multicolumn{7}{l}{\textbf{CausalPFN}}\\
\hspace{1em}\hspace{1em}CausalPFN & 67.6 & \textcolor{red}{\textbf{9.1}} & 73.7 & 19.68 & \textcolor{orange}{\textbf{2.06}} & 9\\
\bottomrule
\end{tabular}}
\begin{tablenotes}%
\tiny
\item CI: confidence interval; GLM: logistic regression without interaction terms; IPTW: inverse probability of treatment weighting.
\item Color legend: \textbf{\textcolor{red}{Red}} = best, \textbf{\textcolor{orange}{Orange}} = second best, \textbf{\textcolor{blue}{Blue}} = third best.
\item *Tied for best result. 
\end{tablenotes}
\end{table}

\FloatBarrier

Across all scenarios and sample sizes, CausalPFN and TabPFN-based single model g-computation consistently produced low $95\%$ CI co\-ve\-rage rates. CausalPFN yielded $95\%$ CI co\-ve\-rage rates of 54.3\% and 67.6\%, respectively, for the scenarios Indomethacin and Rotterdam with sample size ${n=1,000}$. TabPFN-based single model g-computation yielded $95\%$ CI co\-ve\-rage rates of 88.1\% and 86.4\%, respectively, for the scenarios Indomethacin and Rotterdam with sample size ${n=1,000}$. TabPFN-based g-computation with one model per treatment group (T-learner) had higher $95\%$ CI coverage rates: 93.7\% and 91.5\%, respectively, for the scenarios Indomethacin and Rotterdam with sample size ${n=1,000}$. In contrast, TabPFN-based IPTW achieved near-nominal $95\%$ CI co\-ve\-rage rates, especially in the Indomethacin scenario (respectively 94.3\%, 94.7\%, and 95.2\% for sample sizes ${n=200}$, ${n=500}$, and ${n=1,000}$).

When assessing the bias-eliminated $95\%$ CI coverage rate, TabPFN-based approaches aligned closely with the $95\%$ nominal level. This correction was weaker for CausalPFN, with bias-eliminated $95\%$ CI coverage rates below 85\% across all scenarios and sample sizes. CausalPFN and TabPFN-based single model g-computation tended to produce narrower $95\%$ CI than other methods.

\subsection*{Point Estimate Metrics}

\begin{figure}[ht]%
\centering
    \includegraphics[width=0.9\textwidth]{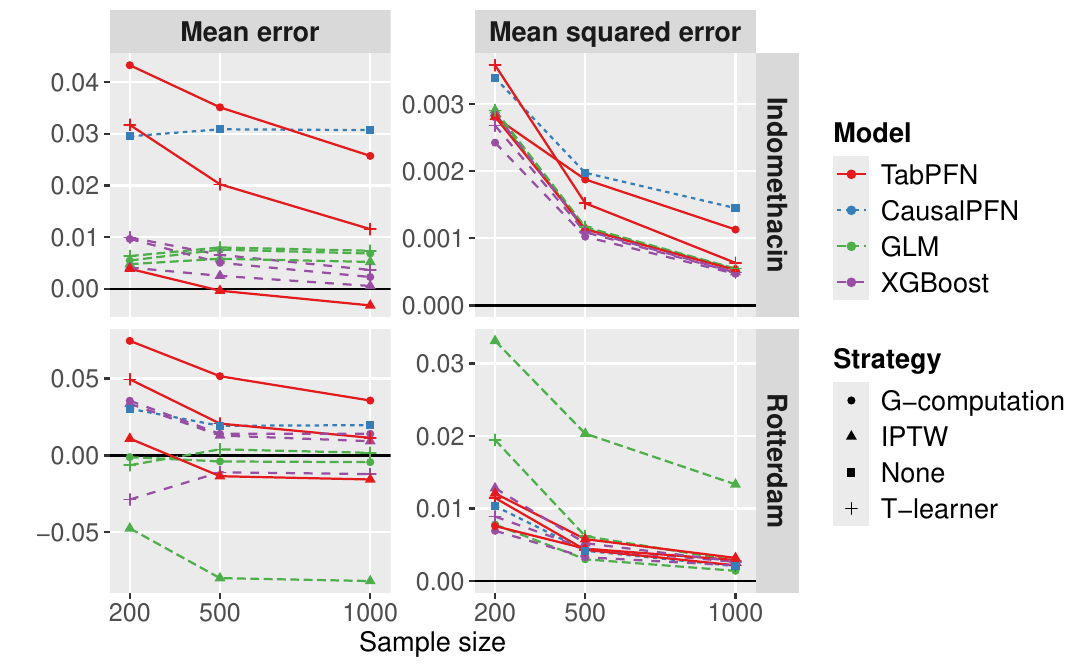}
    \captionof{figure}{Point estimates metrics for the two simulation scenarios. GLM: logistic regression without interaction terms. IPTW: inverse probability of treatment weighting. G-computation corresponds to g-computation with a single model. T-learner corresponds to g-computation with one model per treatment group. The performance of the ``crude association" approach is not shown in this figure.}\label{bias_plots}
\end{figure}

Figure~\ref{bias_plots} shows the results for the point estimate metrics.
Across all scenarios and sample sizes, TabPFN-based g-computation (single model and T-learner) and CausalPFN exhibited large mean errors. On the other hand, TabPFN-based IPTW produced low mean errors. Moreover, CausalPFN yielded large MSE in the Indomethacin scenario.

\subsection*{Computation time}

TabPFN-based methods exhibited computation times to retrieve a point estimation and a confidence interval by bootstrapping with $599$ iterations ranging from ${785}$~s for IPTW in the Indomethacin scenario with ${n=200}$ to ${16,504}$~s for single-model g-computation in the Rotterdam scenario with ${n=1,000}$.
All other procedures, including CausalPFN, exhibited computation times inferior to $42$~s across all scenarios and sample sizes.

\section*{Discussion}\label{discussion}

This study aimed to assess the performance and applicability of PFNs for estimating the ATE in observational clinical research. We found that TabPFN-based ATE estimators using g-computation or IPTW required substantial computation times that increased with sample size, making them impractical for routine use. Moreover, TabPFN-based g-computation with a single model (as opposed to g-computation with one model per treatment group---T-learner) was associated with a large mean error in both simulation scenarios, which translated into inadequately low 95\% CI coverage rates. CausalPFN, which automates causal modeling and thus eliminates the need for g-computation or IPTW, also exhibited inadequately low 95\% CI coverage rates in both simulation scenarios. However, its mean error was large in only one scenario, suggesting that the method CausalPFN uses to construct CIs may contribute to these low coverage rates. This hypothesis is further supported by the fact that correcting CIs for bias resulted in only a modest improvement in coverage for CausalPFN.

These findings are consistent with the well-known fact that, while optimizing prediction error, machine learning algorithms can introduce bias in estimation~\cite{naimi_2023_challenges_valide_causal_effect_ML,Chernozhukov_2018_doubleML_for_trt_and_structural_parameters}. The computation times for TabPFN, however, were surprising. Hollmann et al. reported that TabPFN could be trained in approximately 0.1~s for datasets with $10,000$ observations and 10 features~\cite{hollmann2025tabpfn}, referring to the ICL step (i.e., applying the function $g$). They also noted, however, that making a prediction for a single observation could take around ${0.2}$~s (i.e., applying the function $f_\mathbf{D}$ to one observation). The need for individual predictions in IPTW and g-computation, together with bootstrapping, explains the long computation times we observed.

Our study has several strengths. First, the simulation framework captures complex, real-world–inspired interdependencies between binary variables by incorporating all possible interactions. At the same time, it ensures that the key assumptions of positivity, conditional exchangeability, and consistency are satisfied. Second, while several studies have assessed PFNs for ATE estimation~\cite{zhang2025tabpfnmodelruleall,robertson2025dopfn}, our study is, to the best of our knowledge, the first to evaluate the coverage rates of confidence intervals---a fundamental criterion in estimation. The coverage of CausalPFN's credible intervals was assessed, but without attempting to mimic real-world applications~\cite{balazadeh2025causalpfn}.

Our study has several limitations. The simulations were restricted to two scenarios with binary outcomes, treatments, and confounders---a common but simplified framework in epidemiology. This design allowed us to verify structural positivity while incorporating complex interdependencies, as the degrees of freedom increase exponentially with the number of binary variables. Future work could explore continuous variables and high-dimensional settings. However, parametric assumptions will be necessary to ensure positivity. The binary outcome explains our use of CausalPFN, since Do-PFN is currently limited to continuous outcomes~\cite{robertson2025dopfn}.

The covariates supplied to the models constituted a minimal adjustment set. Future research could investigate the ability of PFNs to identify and exclude irrelevant variables, or to search for an optimal adjustment set that minimizes estimator variance.

We did not examine simulated samples for levels of $Z$ lacking observations from each level of $X$---a phenomenon referred to as ``random nonpositivity”, i.e., random violations of positivity due to finite sample sizes while probability is different from null in the source population~\cite{whatif}. Because positivity was ensured at the population level by the simulation design, such an examination was unnecessary. If ``random nonpositivity" did occur in some simulated samples, the models used in this study were sufficiently inflexible to estimate treatment probabilities in sparse strata using information from individuals in other strata. Likewise, in the IPTW strategy, we did not assess covariate balance after weighting. This was not necessary either, as conditional exchangeability given $Z$ was guaranteed by the simulation design.

In conclusion, the computational burden of TabPFN and the limited performance of CausalPFN currently preclude their practical use in causal inference. Nevertheless, these algorithms are recent developments and may become valuable tools in the future, particularly if trained on real-world data. 
While there is a trend toward automating parts of scientific production~\cite{Musslick2025_automatingthepracticeofscience}, we emphasize that statistical and causal modeling will remain an opportunity to leverage prior information by incorporating domain knowledge into the analysis---a process that is not yet routinely automated.

\section*{Supplementary Materials}
Supplementary material is available below.

\section*{Data and code availability}
Our code and simulations are available at \href{https://github.com/franciscomourao/PFNs_Causal_Inference}{https://github.com/franciscomourao/PFNs\_Causal\_Inference}.

\section*{Competing interests}
No competing interest is declared.

\section*{Author contributions statement}
F.M. and B.G. conceived and conducted the experiment, analyzed the results, and wrote the manuscript. All authors (F.M., D.H., D.B., B.B., N.L., F.C., and B.G.) reviewed and edited the manuscript.

\section*{Acknowledgments}
A CC-BY license has been applied by the authors to this work and will 
be applied to the Author Accepted Manuscript arising from this submission, 
in accordance with the institutions' open access conditions.

\bibliographystyle{ama.bst}
\bibliography{reference.bib}
%\printbibliography

\includepdf[fitpaper=true, pages=-]{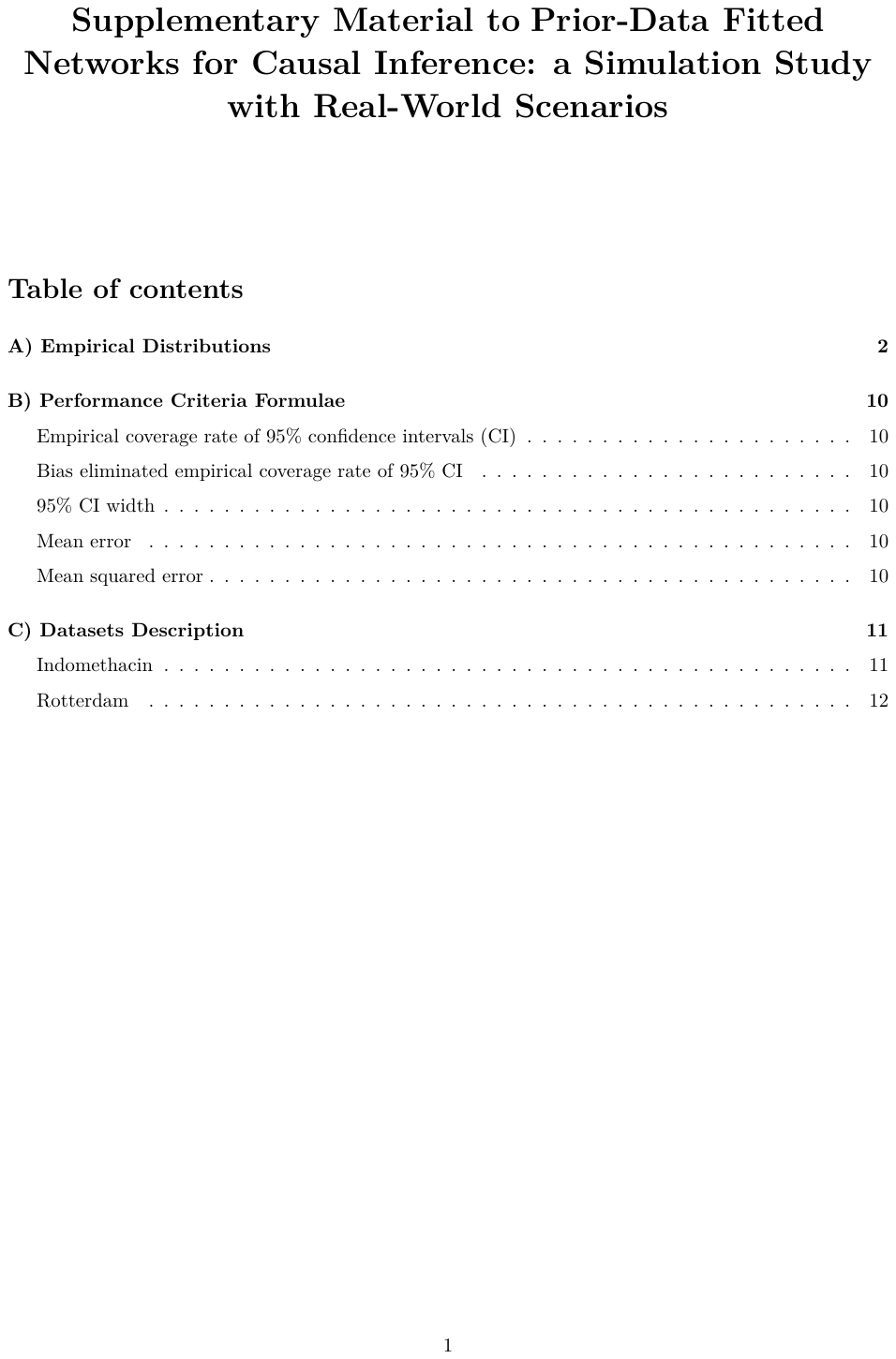}

\end{document}